\documentclass[%
 reprint,
]{revtex4-1}
\usepackage[table]{xcolor}
\usepackage{graphicx,subcaption}
\usepackage{dcolumn}
\usepackage{bm}
\usepackage{amsmath}
\usepackage{amsfonts}
\usepackage{amssymb}
\usepackage{mathrsfs}
\usepackage{graphicx}
\usepackage{color}
\usepackage{hyperref}
\usepackage{multirow}
\usepackage{slashed}
\usepackage{epsf}
\usepackage{mathtools}
\usepackage{amsmath,amssymb}
\usepackage{verbatim}
\usepackage{subcaption}

\usepackage{color}
\usepackage{cancel}
\numberwithin{equation}{section}
\definecolor{nicered}{rgb}{.7,.1,.1}
\definecolor{nicegreen}{rgb}{.1,.75,.15}

\usepackage{amsmath}
\usepackage{amsfonts}
\usepackage{amssymb}
\usepackage{mathrsfs}
\usepackage{graphicx}
\usepackage{color}
\usepackage{hyperref}
\usepackage{multirow}
\usepackage{slashed}
\usepackage{epsf}
\usepackage{mathtools}
\usepackage{amsmath,amssymb}
\usepackage{verbatim}
\usepackage{subcaption}

\usepackage{color}
\usepackage{cancel}
\numberwithin{equation}{section}

\usepackage{amsmath, latexsym, amssymb, graphicx, color, multirow, bm, braket, ascmac, bbm}
\usepackage{hyperref}
\usepackage{relsize}
\usepackage{caption}
\usepackage{subcaption}
\usepackage{dsfont}
\usepackage[table]{xcolor}
\usepackage{tabularx}
\usepackage[T1]{fontenc}
\usepackage{mathtools}
\usepackage[capitalize]{cleveref}
\definecolor{nicered}{rgb}{.7,.1,.1}
\definecolor{nicegreen}{rgb}{.1,.5,.1}
\definecolor{darkblue}{rgb}{0,0,.5}
\hypersetup{colorlinks, citecolor=nicegreen,linkcolor=nicered, urlcolor=darkblue}
\usepackage{multirow}
\usepackage{slashed}
\numberwithin{equation}{section}
\usepackage{verbatim}
\usepackage{cancel}

\usepackage{graphicx}

\def\to{\rightarrow}

\def\tl{\tilde l}

\def\to{\rightarrow}

\def\tl{\tilde l}

\def\tl{\tilde l}

\def\tl{\tilde}

\begin{document}

\preprint{APS/123-QED}

\title{Investigating the GmSUGRA in the MSSM through the long-lived bino NLSP at the HL-LHC}

\author{Wenxing Zhang$^{1}$, Waqas Ahmed$^{2}$, Imtiaz Khan$^{3,4}$, Tianjun Li$^{3,4}$, Shabbar Raza$^{5}$}

\affiliation{\vspace{2mm} \\
	$^1$Tsung-Dao Lee Institute and School of Physics and Astronomy, \\Shanghai
	Jiao Tong University, 800 Dongchuan Road, Shanghai 200240, China. \\
		$^2$School of Mathematics and Physics, Hubei Polytechnic University, Huangshi 435003,China \\
        $^3$CAS Key Laboratory of Theoretical Physics, Institute of Theoretical Physics, \\
            Chinese Academy of Sciences, Beijing 100190, China. \\
        $^4$School of Physical Sciences, University of Chinese Academy of Sciences, No. 19A Yuquan Road, Beijing 100049, China.\\
$^5$Department of Physics, Federal Urdu University of Arts, Science and Technology, Karachi 75300, Pakistan.\\
}

\begin{abstract}

The axino, the supersymmetric partner of axion, is a well-motivated warm/hot dark matter candidate, and provides
a natural solution to the relic density problem for the bino-like neutralino 
if it is the lightest supersymmetric particle (LSP). With the Generalized Minimal Supergravity, we study
such kind of the viable parameter space where the bino-like neutralino is the next-to-LSP (NLSP) and the axino is the LSP.
In addition, we consider a scenario where the bino is a long-lived NLSP with the lifetime varying 
from $10^{-6}$s to $10^{-4}$s, and then propose a new signal searching scheme involving one displaced photon together 
with the large missing transverse momentum at the HL-LHC.
The bino-like lightest neutralino lies under or around 100 GeV and is produced as a decay product of the 
right-handed sleptons.
The relevant axion coupling $f_a$ can be probed up to $\mathcal{O}(10^9)$ GeV at 2$\sigma$ level for the right-handed slepton mass under 300 GeV and the lightest neutralino mass under 100 GeV.

\end{abstract}

\maketitle


\section{\label{sec::introduction} Introduction}

The Penccei-Quinn (PQ) solution~\cite{Peccei:1977hh, Peccei:1977ur} stands out among a variety of solutions to the strong CP problem due to its simplicity and elegance. The theory proposed the existence of a global $U(1)$ symmetry that is spontaneously broken at the PQ scale $f_a$. 
A new Goldstone particle, axion, is therefore predicted.
In low-energy supersymmetry (SUSY) theory relevant with the PQ solution, the axino, as the supersymmetric partners of the axion~\cite{Weinberg:1977ma, Wilczek:1977pj, Kim:1979if, Shifman:1979if, Dine:1981rt, Zhitnitsky:1980tq}, is considered as an alternative lightest supersymmetric particle (LSP) and thus a dark matter (DM) candidate~\cite{Preskill:1982cy, Baer:2014eja}.
Meanwhile, a long-held paradigm in searching for DM components is that DM is mostly composed of cold DM. 
In the theory of Supersymmetry (SUSY), this translates into the weakly interacting massive particles (WIMPs), 
e.g., the lightest neutralino. 
If bino-like neutralino is the LSP, which was the most favored scenario in SUSY models with 
gaugino mass unification at the scale of Grand Unified Theory (GUT), one typically obtains an extremely small annihilation cross section and hence an exceedingly large values of the DM relic density, which usually lies two-to-four orders of magnitude above the measured value~\cite{Roszkowski:2017nbc, Baer:2014eja}. 
Such scenario inspires physicists that the bino-like neutralino is the next-to-lightest supersymmetric particle (NLSP) and the DM is composed of the bino-axino mixture with the axino being the LSP~\cite{Baer:2008yd, Boyarsky:2008xj}, where the overwhelming DM relic density is suppressed by a factor of $m_{\tilde{a}}/m_{\tilde{\chi}_1^0}$~\cite{Covi:2004rb}.

On the other hand, the current Large Hadron Collider (LHC) sets strong constraints on the mass of SUSY particles. For instance, the gluino and squarks are excluded around 2 TeV~\cite{ATLAS:2020syg,CMS:2019zmd}, and the electroweakinos and sleptons are beyond about a few hundred GeV depending on the assumptions~\cite{ATLAS:2019wgx}. Moreover, SUSY models inspired by new measurements involving $g_{\mu}-2$ \cite{Muong-2:2021ojo} indicates  the slepton mass around a few hundred GeV to TeV scale (for example see \cite{Ahmed:2021htr} and references there in).

Among the available scenarios where one can avoid the stringent SUSY search constraints and still remains consistent with ongoing searches is the 
 Electroweak Supersymmetry (EWSUSY) \cite{Cheng:2012np,Cheng:2013hna,Li:2014dna}, where the squarks and/or gluinos are
around a few TeV while the sleptons, sneutrinos, bino and winos are within 1 TeV. The higgsinos (or say the Higgs bilinear $\mu$ term) can be either heavy or light. In particular, the EWSUSY can be realized in the Generalized Minimal Supergravity (GmSUGRA) \cite{Li:2010xr,Balazs:2010ha}. This article is the continuation of our studies of the 
Supersymmetric Standard Models (SSMs) under the light of current and future SUSY searches \cite{Ahmed:2021htr,Ahmed:2022ude}. In particular, in ref.~\cite{Ahmed:2022ude} it is shown that for the $Z$-pole case the right-handed selectron is excluded up to 180 GeV and 210 GeV respectively at 3$\sigma$ and 2$\sigma$, while in case Higgs-pole solutions, the right-handed selectron is excluded up to 140 GeV and 180 GeV respectively at 3$\sigma$ and 2$\sigma$. In this study we consider those solutions which have cold dark matter relic density larger than the Planck2018 5$\sigma$ bounds ~\cite{Planck:2018nkj}. In this study we assume that the lightest neutralino (relatively long-lived) serves as the NLSP which decays to the LSP axino and a photon.

The SUSY particle mass seems to lie beyond the LHC detection ability.
However, SUSY particle searching schemes involving long-lived neutralinos decay are usually implemented through reconstructing a displaced vertex via hard jets~\cite{Dine:1994vc, ATLAS:2018tup, MATHUSLA:2020uve, Lee:2018pag, ATLAS:2018tup}, and a few detection signals involving non-pointing photon, light sleptons and large missing $E_T$ without hard jets induced by decay of light slepton have not been widely investigated~\cite{ATLAS:2022vhr, ATLAS:2014kbb, ATLAS:2022vkf}. 
Hence there still exists possibilities that the bino-like neutralino lies under 100 GeV and the slepton mass around a few hundred GeV~\cite{Ahmed:2022ude}.

In contrast with other SUSY particles, the mass of the axino is model-dependent and remains unconstrained both experimentally and theoretically. For example, in a straightforward SUSY version of the DFSZ model, the axino mass is typically rather small and around $\sim$MeV.
Depending on the model and SUSY breaking scale, the axino mass varies from eV to GeV~\cite{Tamvakis:1982mw, Rajagopal:1990yx, Goto:1991gq, Chun:1992zk}. 
In addition, since the axino coupling to the normal matter is strongly suppressed by a coupling of $1/f_a$~\cite{Raffelt:2006cw}, all heavier SUSY particles cascade-decay to the NLSP in stead of the axino LSP.
Hence it is reasonable to consider the case that the bino decays to axino and a photon ( $\tilde{\chi}_{0}^{1} \to \tilde{a} \gamma$ ) at colliders with the bino being the NLSP whose mass lies under 100 GeV. 
Depending on the magnitude of the PQ symmetry breaking scale $f_a$, such scenario offers a new avenue for exploring SUSY and DM physics:
\begin{itemize}
    \item For $f_a \gtrsim 10^{12}$ GeV, the axinos are produced one second later than the Big Bang. The injection of high energy hadronic and electromagnetic particles~\cite{Freitas:2009fb, Freitas:2009jb} can affect the abundance of light elements produced during Big Bang Nucleosynthesis (BBN)~\cite{Kawasaki:2004yh, Jedamzik:2004er}, such that the BBN constraint can be severe as discussed in Ref.~\cite{Covi:2001nw, Covi:2004rb, Covi:1999ty, Freitas:2011fx}.
    
    \item For $10^{9-10} ~\text{GeV}\lesssim f_a \lesssim 10^{12}$ GeV, there exists possibilities that bino NLSP becomes a long-lived particle~\cite{MATHUSLA:2020uve, No:2019gvl, Alimena:2019zri}. In the tracker, the photon would leave a straight trajectory that does not cross the origin. The signal has very clean background~\cite{ATLAS:2022vhr} and therefore deserves to simulate before the running of the HL-LHC. We focus on this senario in this work.
    \item For $f_a \lesssim 10^{7-8}$ GeV, the bino-like neutralino prompt decays to axino and gamma ray, leaving a signal of two photons plus large missing energy, which background is very clean at the lepton colliders and is widely investigated in Ref.~\cite{Chen:2021omv, ATLAS:2016poa, Brandenburg:2005he, Freitas:2011fx}. By the way,
to evade the cooling constraints from the stars and supernovae, we can assume that the axion
has flavour-violating couplings to the Standard Model (SM) fermions, for example, the axion mainly couples
to the third generation of the SM fermions.

\end{itemize}

It is well-known that there are two kinds of the viable invisible axion models which can satisfy
the experimental bounds: (1) the Kim-Shifman-Vainshtein-Zakharov (KSVZ) axion model, 
which introduces a SM singlet and a pair of extra vector-like quarks that are charged under $U(1)_{PQ}$ 
while the SM fermions and Higgs fields are neutral~\cite{Kim:1979if, Shifman:1979if}; 
(2) the  Dine-Fischler-Srednicki-Zhitnitsky (DFSZ) axion model, in
which a SM singlet and one pair of Higgs doublets are introduced, and the SM fermions and
Higgs fields are all charged under $U(1)_{PQ}$ symmetry~\cite{Dine:1981rt, Zhitnitsky:1980tq}.
{In this paper, we shall consider the DFSZ axion model, and
 a SUSY scenario where the bino is a long-lived NLSP with the lifetime varying 
from $10^{-6}$s to $10^{-4}$s, and then propose a new signal searching scheme involving one displaced photon together 
with the large missing transverse momentum at the HL-LHC}.
The bino-like lightest neutralino lies under or around 100 GeV and is produced as a decay product of the 
right-handed sleptons.
The relevant axion coupling, $f_a$, can be probed up to $\mathcal{O}(10^9)$ GeV at 2$\sigma$ level for the right-handed slepton mass under 300 GeV and the lightest neutralino mass under 100 GeV. 
Also, the KSVZ model can be discussed similarly.

This paper is organized as follows. Section~\ref{sec::model} introduce the framework of the GmSUGRA model.  
Section~\ref{sec::scanning} presents the variable scanning parameter space.
In section~\ref{sec::collider_search}, we discuss the related collider analysis and the kinetic variables. 
At the end of this section, numerical results are presented. Conclusions are present in the last section.

\section{The GmSUGRA Model }\label{sec::model}

Within the framework of the GmSUGRA model, it is possible to implement the EWSUSY. According to this model, the masses of sleptons and electroweakinos (charginos, bino, wino, and/or higgsinos) are all contained inside one TeV. On the other hand, the masses of squarks and/or gluinos may vary across many TeV. Additionally, both the gauge coupling link and the gaugino mass relation are symmetrical at the GUT scale.
\begin{equation}
 \frac{1}{\alpha_2}-\frac{1}{\alpha_3} =
 k~\left(\frac{1}{\alpha_1} - \frac{1}{\alpha_3}\right)~,
\end{equation}
\begin{equation}
 \frac{M_2}{\alpha_2}-\frac{M_3}{\alpha_3} =
 k~\left(\frac{M_1}{\alpha_1} - \frac{M_3}{\alpha_3}\right)~,
\end{equation}
With $k=5/3$, we have a straightforward gaugino mass relation. By assuming gauge coupling unification at the GUT scale ($\alpha_1=\alpha_2=\alpha_3$) we have gauginos relation
\begin{equation}
 M_2-M_3 = \frac{5}{3}~(M_1-M_3)~.
\label{M3a}
\end{equation}
In this case, there are just two independent gauginos rather than three. The expression for $M_3$ may be written in terms of $M_2$ and $M_3$ as follows:
\begin{eqnarray}
M_3=\frac{5}{2}~M_1-\frac{3}{2}~M_2~,
\label{M3}
\end{eqnarray}
Whereas the value of $M_3$ may be as low as few hundred GeV or as high as several TeV, depending on the particular values of $M_1$ and $M_2$. The GUT-scale masses of general SUSY breaking scalars are tabulated in Ref.~\cite{Balazs:2010ha}. The following squark masses are obtained in the SU(5) model with an adjoint Higgs field, where the slepton masses are treated as free parameters.
\begin{eqnarray}
m_{\tl{Q}_i}^2 &=& \frac{5}{6} (m_0^{U})^2 +  \frac{1}{6} m_{\tl{E}_i^c}^2~,\\
m_{\tl{U}_i^c}^2 &=& \frac{5}{3}(m_0^{U})^2 -\frac{2}{3} m_{\tl{E}_i^c}^2~,\\
m_{\tl{D}_i^c}^2 &=& \frac{5}{3}(m_0^{U})^2 -\frac{2}{3} m_{\tl{L}_i}^2~,
\label{squarks_masses}
\end{eqnarray}
where $m_{\tl Q}$, $m_{\tl U^c}$, $m_{\tl D^c}$, $m_{\tl L}$, and  $m_{\tl E^c}$ represents the left-handed squark doublets, right-handed up-type squarks, right-handed down-type squarks, left-handed sleptons, and right-handed sleptons, respectively, and $m_0^U$  is the universal scalar mass, as in the mSUGRA. The light sleptons occur from EWSUSY , $m_{\tl L}$ and $m_{\tl E^c}$ being within 1 TeV. In particular, in the limit $m_0^U \gg m_{\tl L/\tl E^c}$,  we get the estimated relations for squark masses: $2 m_{\tl Q}^2 \sim m_{\tl U^c}^2 \sim m_{\tl D^c}^2$. In addition, the Higgs soft masses $m_{\tl H_u}$ and $m_{\tl H_d}$, and the  trilinear soft terms $A_U$, $A_D$ and $A_E$ can all be free parameters from the GmSUGRA.

\section{Scanning Process}\label{sec::scanning}

In order to carry out random scans over the parameter space described below, we make use of the ISAJET-7.84 software package. Through the use of the MSSM renormalization group equations (RGEs) in the $\overline{DR}$  regularization scheme, the weak scale values of the gauge and third generation Yukawa couplings are evolved in this package to the value  $M_{\rm GUT}$ .
We do not strictly enforce the unification condition $g_3=g_1=g_2$ at $M_{\rm GUT}$ since a few percent departure from unification may be allocated to the unknown GUT-scale threshold corrections. All of the SSB parameters, together with the gauge and Yukawa couplings, are evolved back to the weak scale $M_{\rm Z}$ when the boundary conditions are specified as $M_{\rm GUT}$.  See \cite{Baer:1999sp} for more details on the workings of ISAJET.

Using parameters given in Section~\ref{sec::model}, we  perform
the random scans for the following parameter ranges
\begin{align} \label{input_param_range}
100 \, \rm{GeV} \leq & m_0^{U}  \leq 5000 \, \rm{GeV}  ~,~\nonumber \\
80 \, \rm{GeV} \leq & M_1  \leq 400 \, \rm{GeV} ~,~\nonumber \\
600\, \rm{GeV} \leq & M_2   \leq 1200 \, \rm{GeV} ~,~\nonumber \\
600 \, \rm{GeV} \leq & m_{\tilde L}  \leq 1200 \, \rm{GeV} ~,~\nonumber \\
100 \, \rm{GeV} \leq & m_{\tilde E^c}  \leq 350 \, \rm{GeV} ~,~\nonumber \\
100 \, \rm{GeV} \leq & m_{\tilde H_{u,d}} \leq 5000 \, \rm{GeV} ~,~\nonumber \\
-6000 \, \rm{GeV} \leq & A_{U}=A_{D} \leq 6000 \, \rm{GeV} ~,~\nonumber \\
-600 \, \rm{GeV} \leq & A_{E} \leq 600 \, \rm{GeV} ~,~\nonumber \\
2\leq & \tan\beta  \leq 60~.~
\end{align}
We have also considered $\mu \ge$0 and used $m_{t}=$173.3 GeV. All the data gathered is REWSB compatible, with the neutralino serving as the LSP. To properly interpret the results we need the following constraints (motivated by the LEP2 experiment) on sparticle masses.

\textbf{\underline{LEP constraints:}}
We impose the bounds that the LEP2 experiments set on charged sparticle masses ($\gtrsim 100$ GeV) \cite{ParticleDataGroup:2016lqr}.

\textbf{\underline{Higgs Boson mass:}}
The experimental combination for the Higgs mass reported by 
the ATLAS and CMS Collaborations is \cite{ATLAS:2016neq}
\begin{align}\label{eqn:mh}
m_{h} = 125.09 \pm 0.21(\rm stat.) \pm 0.11(\rm syst.)~GeV .
\end{align}  
Due to the theoretical uncertainty in the Higgs mass calculations in the MSSM -- see {\it e.g.}~\cite{Slavich:2020zjv,Allanach:2004rh} -- we apply the constraint from the Higgs boson mass to our results as: 
\begin{align}\label{eqn:higgsMassLHC}
122~ {\rm GeV} \leq m_h \leq 128~ {\rm GeV}. 
\end{align}

\textbf{\underline{Rare B-meson decays:}} Since the SM predictions are in a good agreement with the experimental results for the rare decays of $B-$meson such as the $B_{s}\rightarrow \mu^{+}\mu^{-}$, $B_{s}\rightarrow X_{s}\gamma$, where $X_{s}$ is an appropriate state including a strange quark, the results of our analyses are required to be consistent with the measurements for such processes. Thus we employ the following constraints from B-physics \cite{CMS:2014xfa,HFLAV:2014fzu}:

\begin{align}\label{eqn:Bphysics}
1.6\times 10^{-9} \leq ~ {\rm BR}(B_s \rightarrow \mu^+ \mu^-) ~
\leq 4.2 \times10^{-9} ,\\ 
2.99 \times 10^{-4} \leq  ~ {\rm BR}(b \rightarrow s \gamma) ~
\leq 3.87 \times 10^{-4}, \\
0.70\times 10^{-4} \leq ~ {\rm BR}(B_u\rightarrow\tau \nu_{\tau})~
\leq 1.5 \times 10^{-4}.
\end{align}

\textbf{\underline{Current LHC searches:}}
Based on \cite{ATLAS:2017mjy,Vami:2019slp,CMS:2017okm}, we consider the following constraints on gluino and first/second generation squark masses
\begin{align}
(a) \quad \quad m_{\widetilde g} \gtrsim ~ 2.2 ~ {\rm TeV},\quad \quad m_{\widetilde q} \gtrsim ~ 2 ~ {\rm TeV},
\label{lhc-a}
\end{align}

\textbf{\underline{DM searches and relic density:}} We apply the following limit for the neutralino relic density in order to facilitate the discussion on the phenomenology of the axino DM in our scenario:
\begin{align}\label{eq:omega}
\Omega_{\tilde \chi_{1}^{0}}h^2 \gtrsim 0.126.
\end{align}

\begin{figure*}[htbp]\centering
\includegraphics[width=0.4\textwidth]{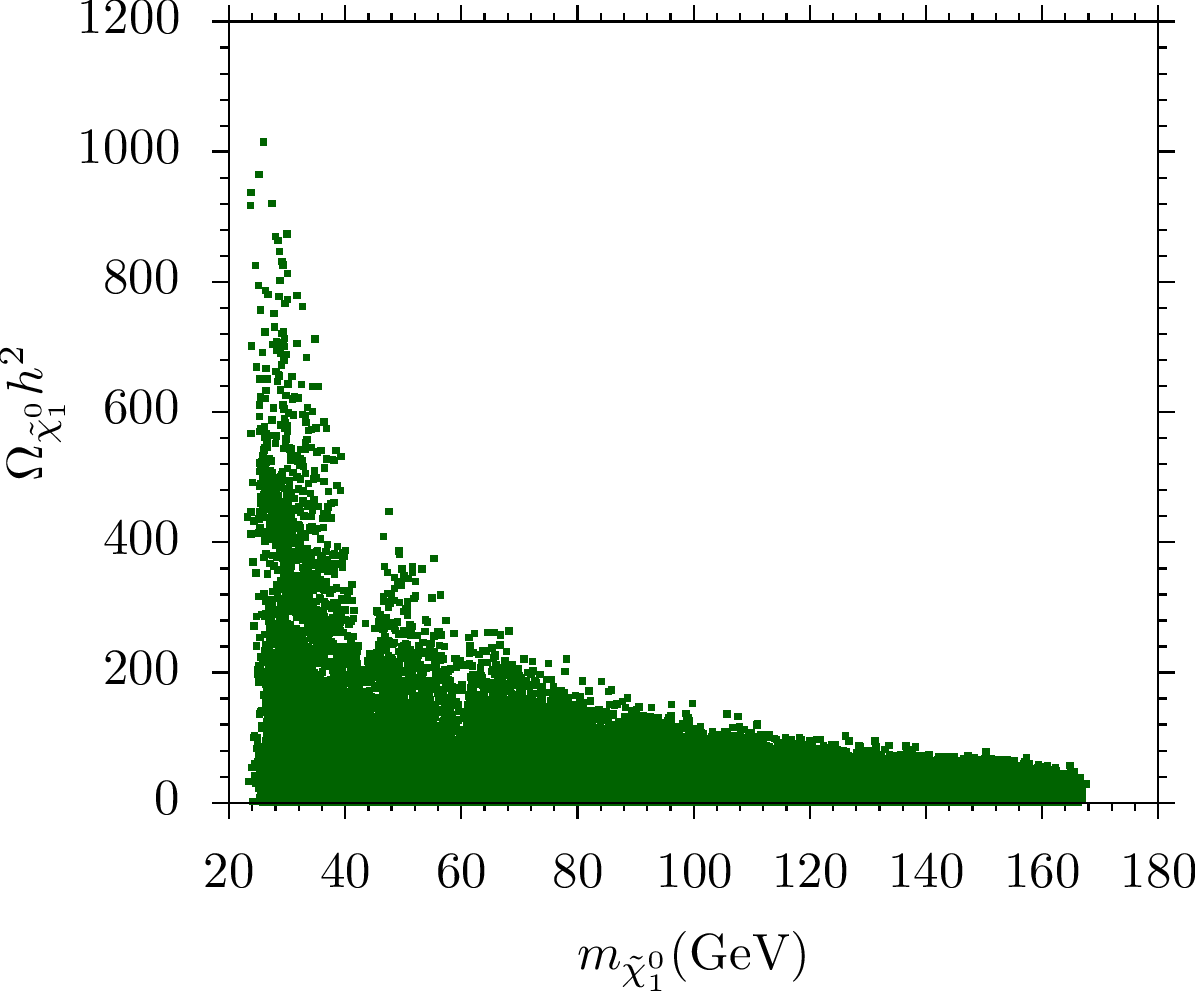}
\includegraphics[width=0.4\textwidth]{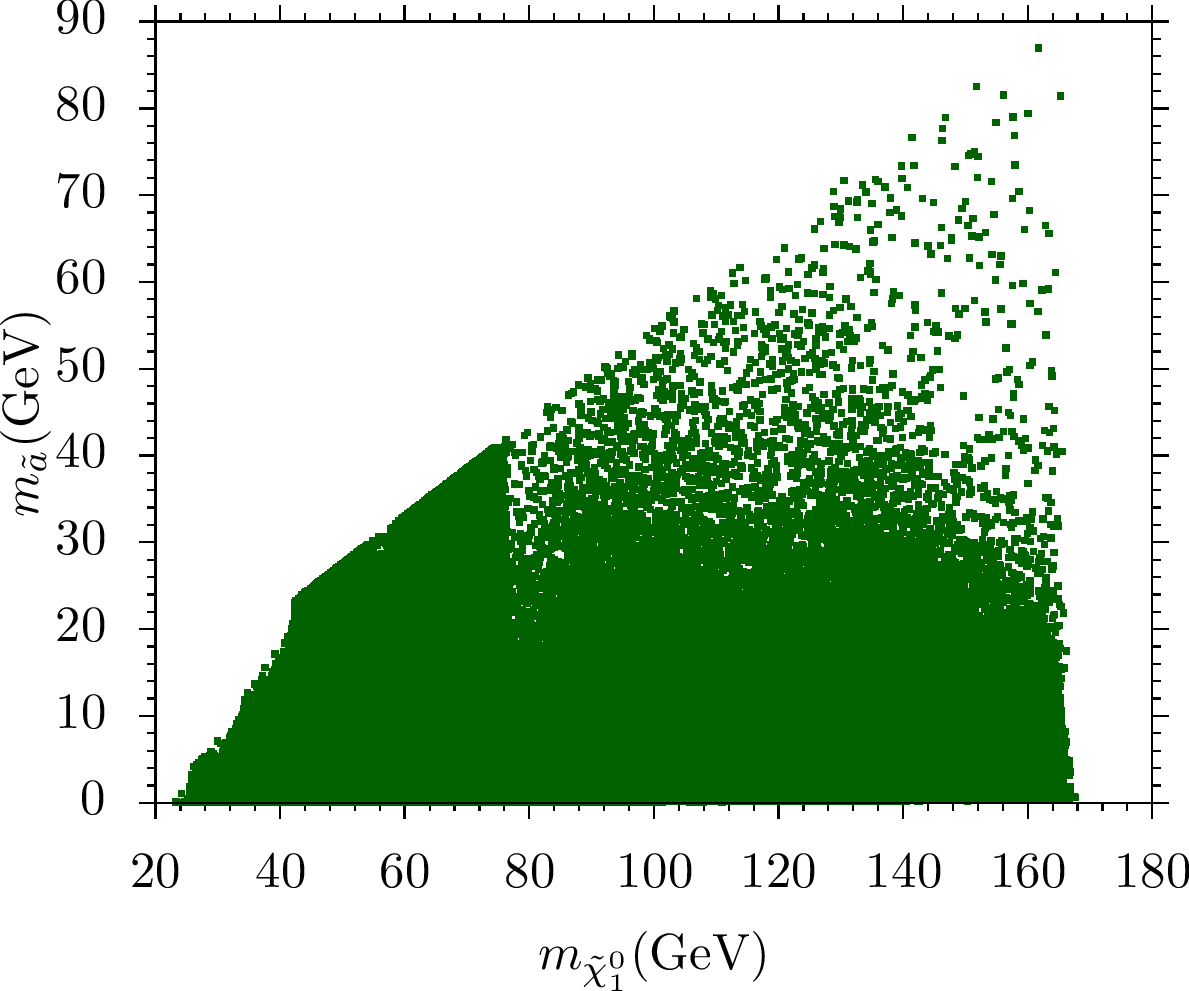}	
	\caption {Plots in $m_{\tilde \chi_{1}^{0}}-\Omega_{\tilde \chi_{1}^{0}} h^{2}$ (left) and $m_{\tilde \chi_{1}^{0}}-m_{\tilde a}$ planes. All the points satisfy REWSB bounds, particle mass bounds, B-physics bounds and $\Omega_{\tilde \chi_{1}^{0}}\gtrsim 0.126$ as described in section \ref{sec::scanning}.}
\label{fig1}
\end{figure*}


\section{\label{sec::scanning} Scans Results}
In this section we show results of our scans only for the relevant parameters. In Fig.~\ref{fig1}, plot in the left panel neutralino relic density ($\Omega_{\tilde \chi_{1}^{0}} h^{2}$) as function of NLSP neutralino mass. Here we have made sure that $\Omega_{\tilde \chi_{1}^{0}} h^{2} \gtrsim$ 0.126 (greater than the Planck2018 5$\sigma$ bounds). It is evident from the plot that in our scans $\Omega_{\tilde \chi_{1}^{0}} h^{2}$ can be as large as 1000 while $m_{\tilde \chi_{1}^{0}}$ is between 20 GeV to 170 GeV. In the right panel we show plot in $m_{\tilde \chi_{1}^{0}}-m_{\tilde a}$ plane. We calculate the axino mass using the relation $m_{\tilde a}\approx \frac{\Omega_{\tilde a} h^{2}}{\Omega_{\tilde \chi_{1}^{0}} h^{2}}  m_{\tilde \chi_{1}^{0}}$ \cite{Covi:1999ty}.Here we assume that the axino saturates the dark matter relic density bounds with $\Omega_{\tilde a}=$0.11. We see that in our scans, axino mass can be very light but can be as heavy as 90 GeV. We then use the $m_{\tilde \chi_{1}^{0}}$ and $m_{\tilde a}$ to calculate the width $\Gamma(\tilde{\chi}_1^0 \to \tilde{a} \gamma )$ as shown in the next section.

\section{\label{sec::collider_search} Long-lived axino searches at the LHC}
\begin{figure}[htbp]
    \centering
    \includegraphics[width=0.43\textwidth]{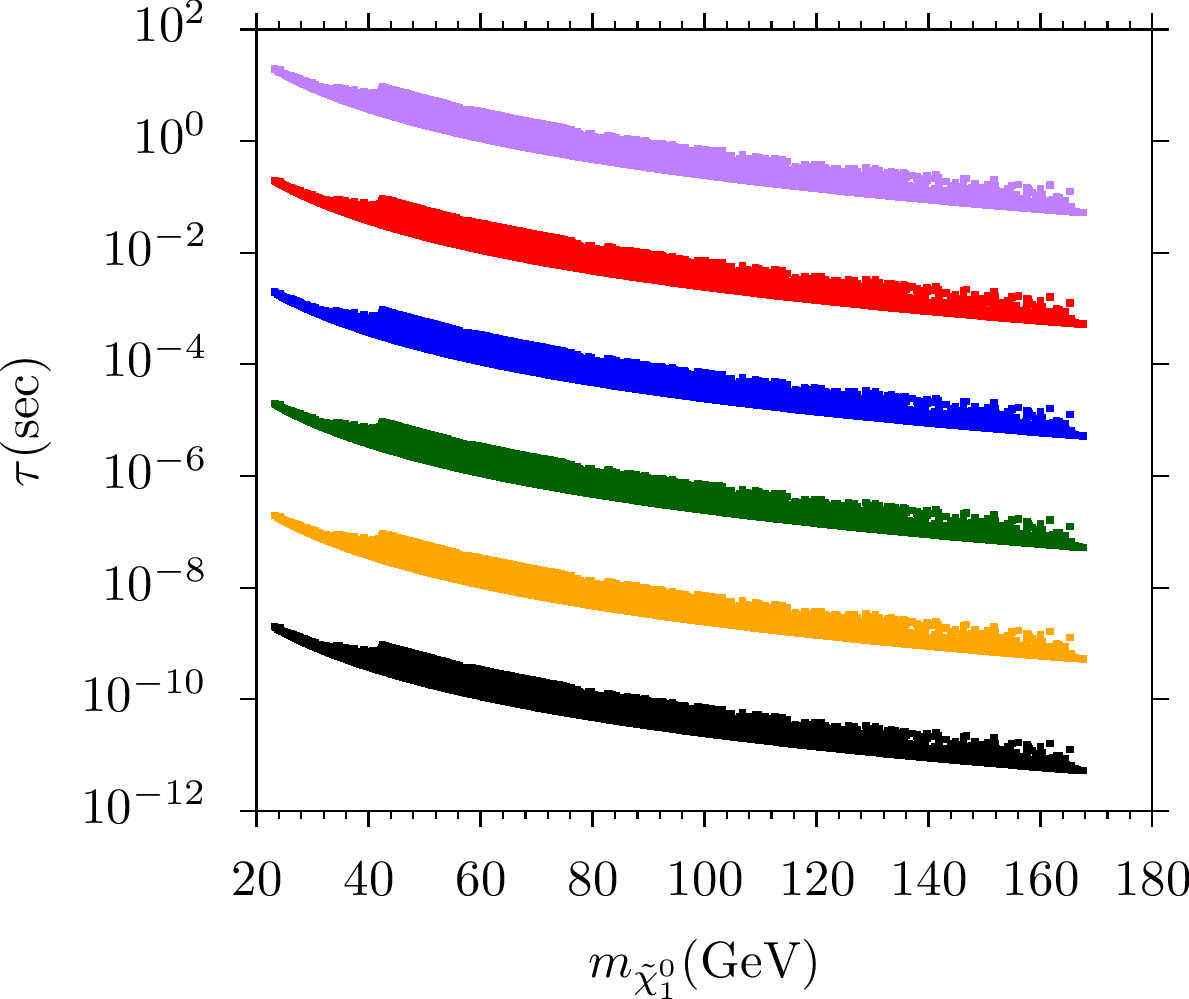}
    \caption{Lifetimes in seconds of the lightest neutralino versus its mass for $f_a/N = 10{^7}, 10^{8}, 10^{9}, 10^{10}, 10^{11}$ and  $10^{12}$ GeV from the bottom to the top.}
    \label{fig:lift_time}
\end{figure}

As the LSP, the axino can be produced by the decay of the unstable NLSP, the lightest neutralino. 
The width $\Gamma(\tilde{\chi}_1^0 \to \tilde{a} \gamma )$ has been calculated in Refs.~\cite{Covi:1999ty, Covi:2001nw}, and is given by
\begin{equation}\label{eq::dw}
    \Gamma(\tilde{\chi}_1^0 \to \tilde{a} \gamma ) = \frac{\alpha^2_{em} C_{aYY}^2 v_4^{(1)2}}{128 \pi^3 \cos^2\theta_W} \frac{m^3_{\tilde{\chi}_1^0}}{(f_a/N)^2} \left( 1-\frac{m^2_{\tilde{a}}}{m_{\tilde{\chi}_1^0}^2} \right),
\end{equation}
where $v_4^{(1)}$ denotes the bino fraction of neutralino of $\tilde{\chi}^0_1$, N is the model-dependent anomaly factor of order $\mathcal{O}(1)$, and $C_{aYY}$ (e.g. $C_{aYY}=8/3$ in the DFSZ model) is a model-dependent coupling factor.
{In the following text, without loss of generality, we consider the DFSZ model as an example. One can check the average decay length of the lightest neutralino lies within 1m when $f_a/N \lesssim 10^7$ GeV. 
However, even if the $f_a/N \sim 10^8$ and $10^9$ GeV, there still exists possibility that the lightest neutralino decay within the Electromagnetic Calorimeter (ECAL), and thus is able to be detected. 
On the other hand, in the case of $f_a/N \gtrsim 10^8$ and $10^9$ GeV, events would be more likely to escape from the current constraints of LHC. Hence the parameter space with $f_a/N \sim 10^{8}$ and $10^{9}$ GeV is essential in digging the long-lived neutralino signal.
In fact, our simulation has shown the parameter region with $f_a/N \lesssim 10^7$ GeV has been excluded by the LHC searches. 
In Fig.~\ref{fig:lift_time},  we plot the $\tilde{\chi}_1^0$ lifetime in seconds versus $m_{\tilde{\chi}_1^0}$ for six choices of $f_a/N$, and taking $C_{aYY}=8/3$ in the DFSZ model.
}

The NLSP decay length depends on the PQ symmetry breaking scale $f_a$, and then provides important information about the $U(1)$ symmetry breaking mechanism.
The very weak coupling of the NLSP to the axino could lead to the displaced vertices (DV) of the NLSP.
In this case, photons are generated in the point of DV, and they reach the ECAL up to $\mathcal{O}(1)$ns later than particles generated in the primary vertex. 
Thus, measuring the photon time of arrival delay with respect to a photon produced at the primary vertex and travailing at the speed of light helps to discriminate between signal and background.
Since the best time resolution for ECAL is measured to be between 70 ps to 100 ps~\cite{CMS:2019zxa}, the ECAL is able to detect the delayed arrived photons.

In our model, at the LHC, the axino can be generated by the cascade decay of the right-handed (RH) sleptons,
 and the concomitant decay product, the photon, is able to be distinguished by the ECAL due to the delayed arrival time.
 
At the HL-LHC, a characteristic signal process associated with the generation of long-lived neutralino is $p ~p \to \tilde{l}_R^- \tilde{l}_R^+ \to l_R^+ l_R^- \tilde{\chi}^0_1 \tilde{\chi}^0_1$ with the slepton being an on-shell particle.
{In the simulation, we generate the signal process and calculate the possibility for the NLSP, $\tilde{\chi}_1^0$, to decay inside ECAL}. 

\begin{figure}[htbp]
    \centering
    \includegraphics[width=0.3\textwidth]{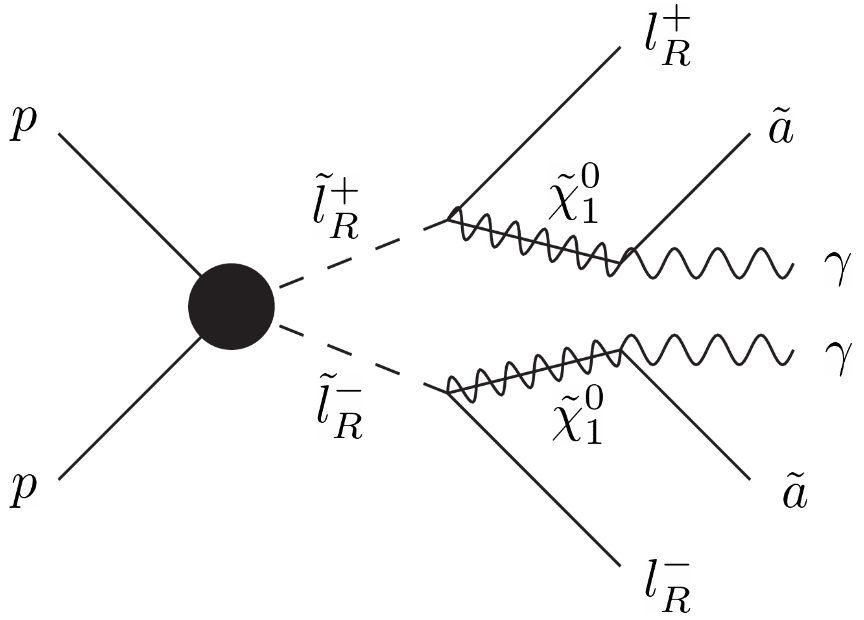}
    \caption{A characteristic signal process associated with the generation of the long-lived neutralino.}
    \label{fig:cross_section}
\end{figure}

The geometrical acceptance probability for the LSP with a decay length $d$ as it traverses the detector is given by
\begin{equation}\label{eq::possibility}
    p = \frac{1}{4\pi} \int_{\Delta\Omega} d\Omega \int_{L_1}^{L_2} \frac{1}{d} e^{-L/d},
\end{equation}
where $L_1$ and $L_2$ are the distances between the interaction point to the point where the Long Lived Particle (LLP) enters and exits the decay volume, and $\Delta\Omega$ is the cross section of the active detector volume~\cite{Curtin:2018mvb, Banerjee:2019ktv}. 
In our simulation, $L_1$ is chosen to be the interaction point, $L_2$ is the point for $\tilde{\chi}_1^0$ to exist the ECAL, and $\Delta\Omega$ is the solid angle covered by the ECAL barrel.

\begin{figure}[htbp]
    \centering
    \includegraphics[width=0.45\textwidth]{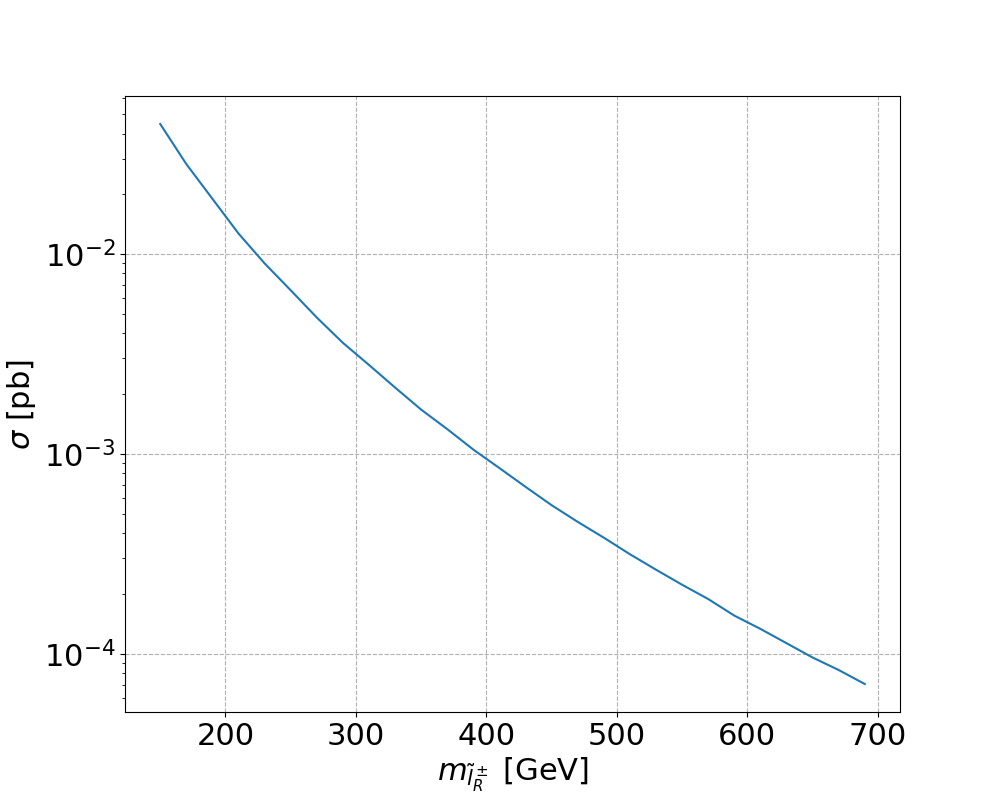}
    \caption{The axino  $\tilde{a}$ production cross section at  the HL-LHC.}
    \label{fig:cross_section}
\end{figure}

As for the photons, they can be detected by the barrel region of the ECAL detector ($|\eta|<1.444$) with its mother particle $p_T > 70$ GeV are labeled as tight photon~\cite{CMS:2019zxa}. 
On the other hand, the photons, which can be detected by the whole region of the ECAL detector ($|\eta|<2.37$) with its mother particle $p_T > 50$ GeV, are labeled as loose photon~\cite{ATLAS:2014kbb}. 
Both photons are required to be isolated, by requiring that the transverse energy  deposited in the calorimeter in a cone of radius $\Delta R=\sqrt{(\Delta\eta)^2+(\Delta\phi)^2}=0.4$.
 
The Monte Carlo samples of signal events are generated by using the MadGraph5~\cite{Alwall:2011uj} for hard scattering processes, PYTHIA8.2~\cite{Sjostrand:2014zea} for parton showering and hadronization, and DELPHES 3~\cite{deFavereau:2013fsa} for jet clustering and detector simulation.
We generate signal events with collider collision energy being 14 TeV and the luminosity set to be 3000$fb^{-1}$. 
In Fig.~\ref{fig:cross_section}, we plot the axino production cross section in pb versus $m_{\tilde{l}_R}$ at the HL-LHC with the lightest neutralino set to be $m_{\tilde{\chi}_1^0}=50$ GeV. Since the slepton decays to $\tilde{\chi}_1^0$ by 100\%, the axino production cross section is independent of $m_{\tilde{\chi}_1^0}$ at tree level.
In addition, the signal processes with $m_{\tilde{\chi}_1^0} > 500$ GeV generate event number less than $\mathcal{O}(1)$ by 3000$fb^{-1}$ luminosity, and thus induce a considerable statistical error. 
Hence in simulation, we consider the LSP mass ranging from 50 GeV to 130 GeV in steps of 10 GeV, and the right-handed slepton mass from 150 GeV to 500 GeV in steps of 20 GeV.
The signal events with two tight photons are selected out.
With the above calculation of LLP decay possibility, we can then estimate 95\% confidence level (C.L.) exclusion limits under the assumption of zero background, in the RH slepton and neutralino mass planes.

\begin{table}[h!]
	\centering
	\scalebox{0.8}{
		\begin{tabular}{lccccc}
			\hline
			\hline
			& Point 1 & Point 2   &point 3 & point 4 & point 5 \\
			\hline
			$m_{0}^{U}$      &   1927     & 1285       & 2496         & 2111  & 1502    \\
			$M_{1} $         &   116.1    & 238.5      & 153.9        & 186.9 & 220..3\\
			$M_{2} $         &   -814.6   & -635.5     &-1148         & -1014 & -872.9 \\
			$M_{3} $         &  1512.1    & 1549.5     & 2106.8       & 1988.2  & 1860.1\\
			$m_{E^c},m_{L}$  & 292.6,1121 & 154.1,883.8 & 114.8,670.5 & 207.5, 1041 & 294.9,713.2  \\
			$m_{Q}$      & 1763.2 & 1174.7 & 2279   & 1928.9 & 1376.4\\
			$m_{U^{c}}$  & 2476.2 & 1654.1 & 3221   & 2720   & 1924.1\\
			$m_{D^{c}}$  & 2487.7 & 1658.9 & 3222.3 & 2725.3 & 1939.1  \\
			$m_{H_{u}},m_{H_{d}}$  & 1025,1286 & 183.7, 1442 & 2653,1551 & 2325,1566 & 3807, 3069\\
		
			$A_{t}=A_{b},A_{\tau}$ &-5974,61.96 & -5944,-365.3  &-2459,-126.8 & -5380, 260.7 &-2787,559.5\\
			$\tan\beta$  & 6.77 & 44.1  & 41.7 & 11.8 & 22\\
			\hline
			$m_h$            &  123    &125  &122 & 122 &125   \\
			$m_H$            &  3437    &2133   &3414  &3839 &4096  \\
			$m_{A} $         &  3415     &2120   &  3392 &3814 & 4069    \\
			$m_{H^{\pm}}$    &  3438    &   2135   & 3415 & 3840 &4097  \\
			\hline
			$m_{\tilde{\chi}^0_{1,2}}$
			& 44,725 & 98,569&  55, 1015&71,905 & 87,775\\
			$m_{\tilde{\chi}^0_{3,4}}$
			& 3234,-3234 & 2895, 2896 &  -3009,3010 &3101,3101  & 2108,2108 \\
			$m_{\tilde{\chi}^{\pm}_{1,2}}$
			&729,3247  &571, 2900  & 1017,3009 &910, 3106 & 779, 2113 \\
			\hline
			$m_{\tilde{g}}$  & 3320    & 3344  & 4517 & 4262 & 3966   \\
			\hline 
			$m_{ \tilde{u}_{L,R}}$
			& 3349,3685  &  3125,3297   & 4481,4903 &4129,4448 &3852, 3701\\
			$m_{\tilde{t}_{1,2}}$
			& 2230,2705 & 1781, 2226  & 3786,4028 &3324, 3617  & 2212, 2877  \\
			\hline 
			$m_{ \tilde{d}_{L,R}}$
			& 3450,3773 & 3127,3335 & 4481,5013 &4130,4550 &  3701, 3940\\
			$m_{\tilde{b}_{1,2}}$
			& 2672,3730 & 2174,2837 & 3828,4700 & 3579,4487 & 2870, 3708\\
			\hline
			$m_{\tilde{\nu}_{1}}$
			& 1098       & 907    & 645&1021 & 700 \\
			$m_{\tilde{\nu}_{3}}$
			& 1096      &   924   &391 &1006 & 549 \\
			\hline
			
			$m_{ \tilde{e}_{L,R}}$
			& 1106,816  &914, 492  & 635,1032&1029 & 724,810 \\
			$m_{\tilde{\tau}_{1,2}}$
			& 814,1102    &  454, 947 & 240,740 &900, 1020 & 485, 616  \\
			\hline
            $f_{a}$ 
            & $10^{8}$ & $10^{8}$ & $10^{9}$ & $10^{10}$ & $10^{11}$ \\ 
            $m_{\tilde{a}}$
			& 0.055 & 1.995  & 1.993 &0.09 & 0.652\\
			$\Omega_{\tilde{\chi}}h^2$
		& 86.855 &5.38 &  3.044& 87.25 & 14.69 \\
		$\Gamma(\tilde{\chi}_1^0 \to \tilde{a} \gamma )$
			& 2.22$\times 10^{-17}$ & 2.49 $\times 10^{-16}$ & 4.46 $\times 10^{-19}$  &9.65 $\times 10^{-21}$ & 1.8 $\times 10^{-22}$\\
			$LT$
			& 2.97 $\times 10^{-8}$ &2.65$\times10^{-9}$   &1.47 $\times 10^{-6}$ & 6.82 $\times 10^{-5}$ & 3.74 $\times 10^{-3}$
    \\
			\hline
			\hline
		\end{tabular}
	}
	\caption{The PQ symmetry breaking scales ($f_a$), decay width ($\Gamma$), sparticle and Higgs masses in GeV, as well as lifetimes (LT) in seconds.}
		\label{table1}
\end{table}

\begin{figure*}[htbp]\centering
       \centering
\includegraphics[width=0.45\textwidth]{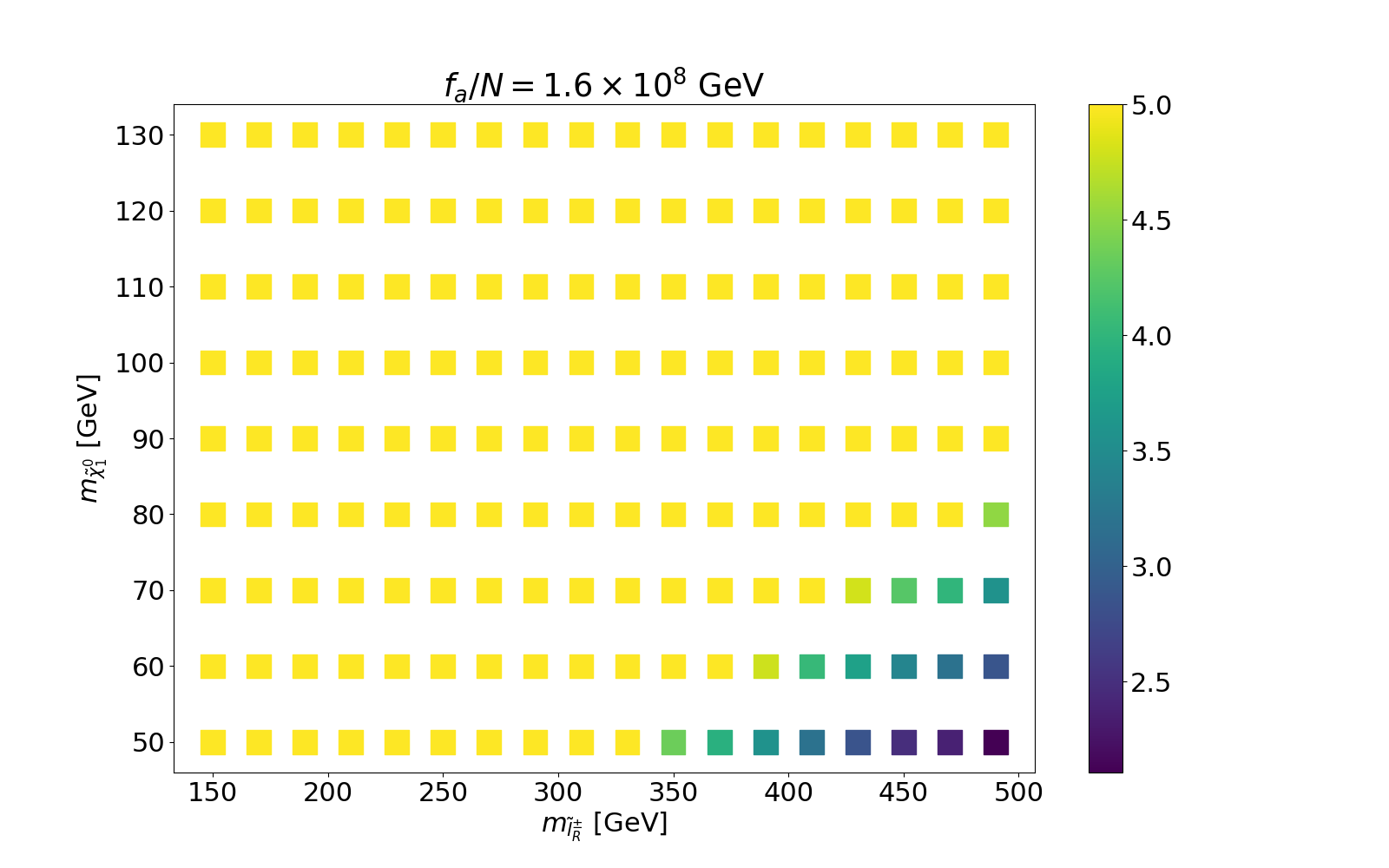}
\includegraphics[width=0.45\textwidth]{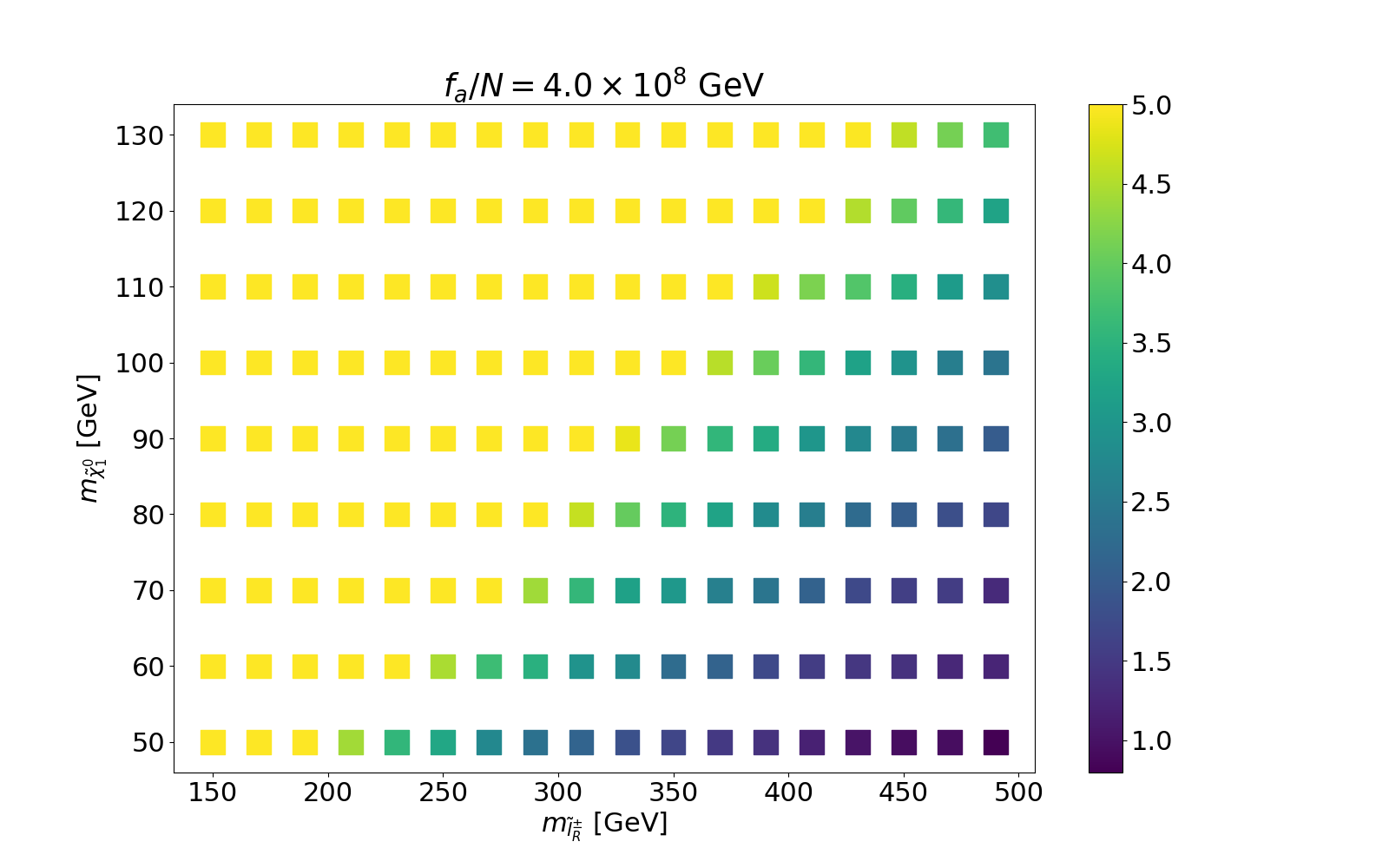}
\includegraphics[width=0.45\textwidth]{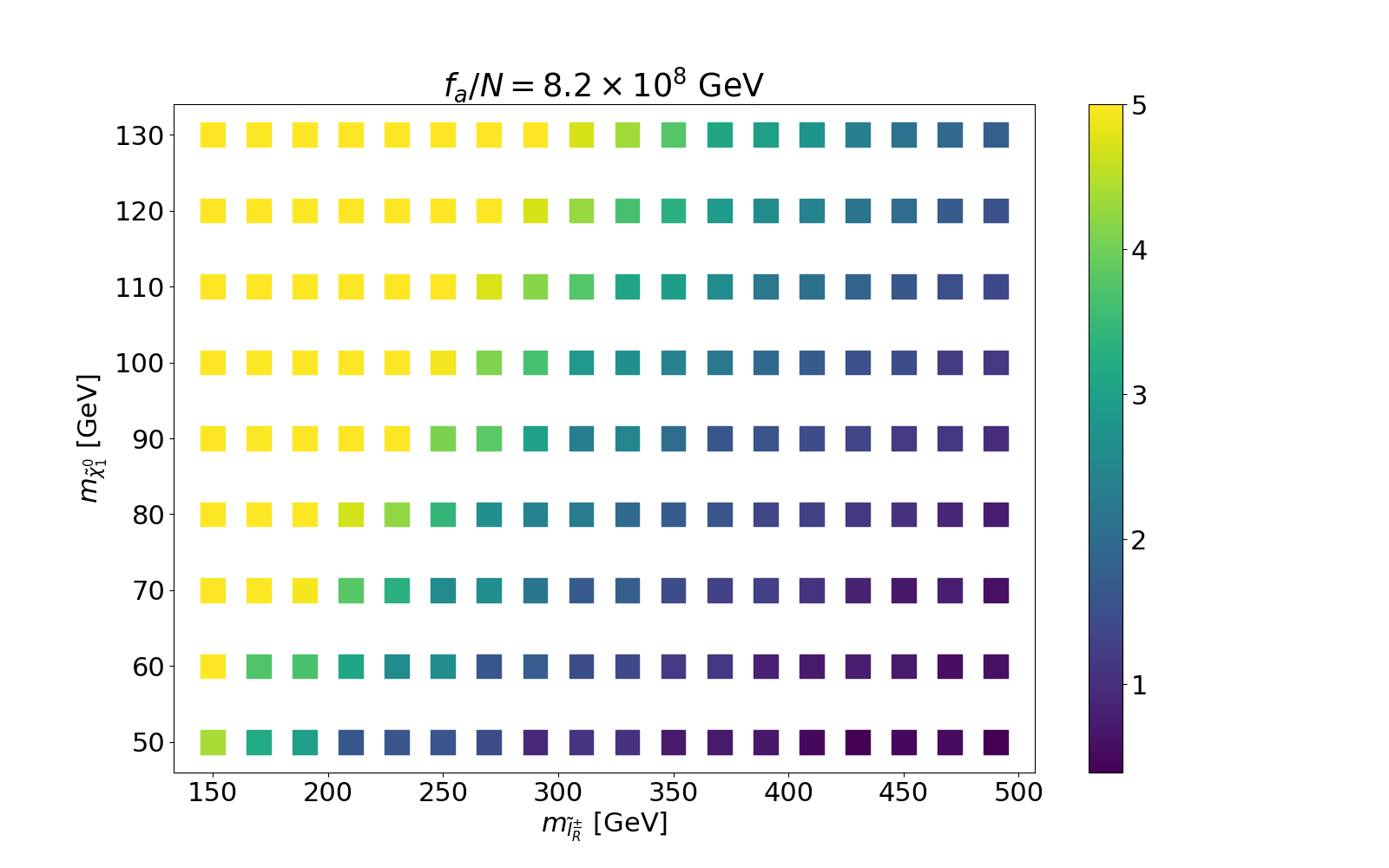}
\includegraphics[width=0.45\textwidth]{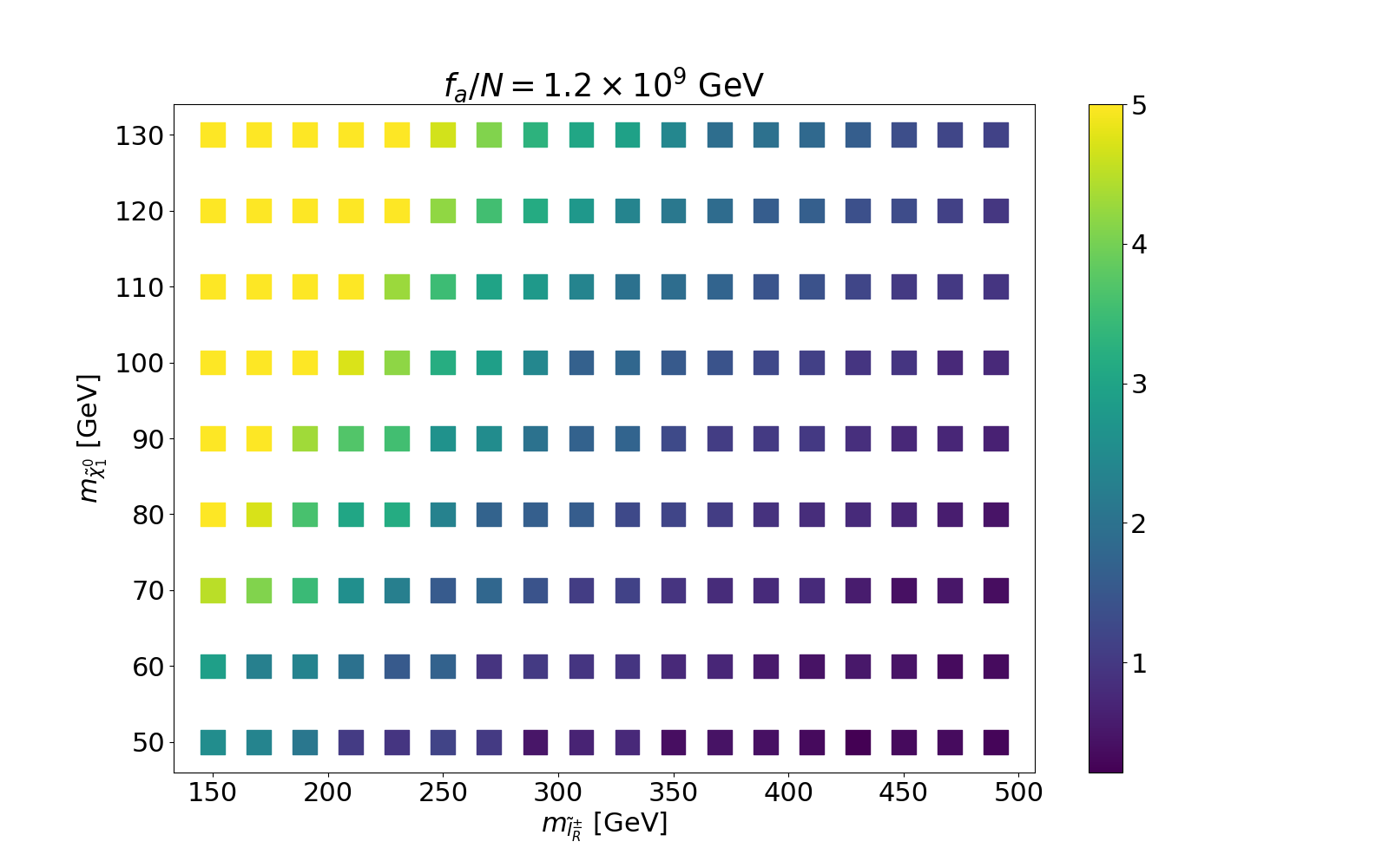}   
\includegraphics[width=0.45\textwidth]{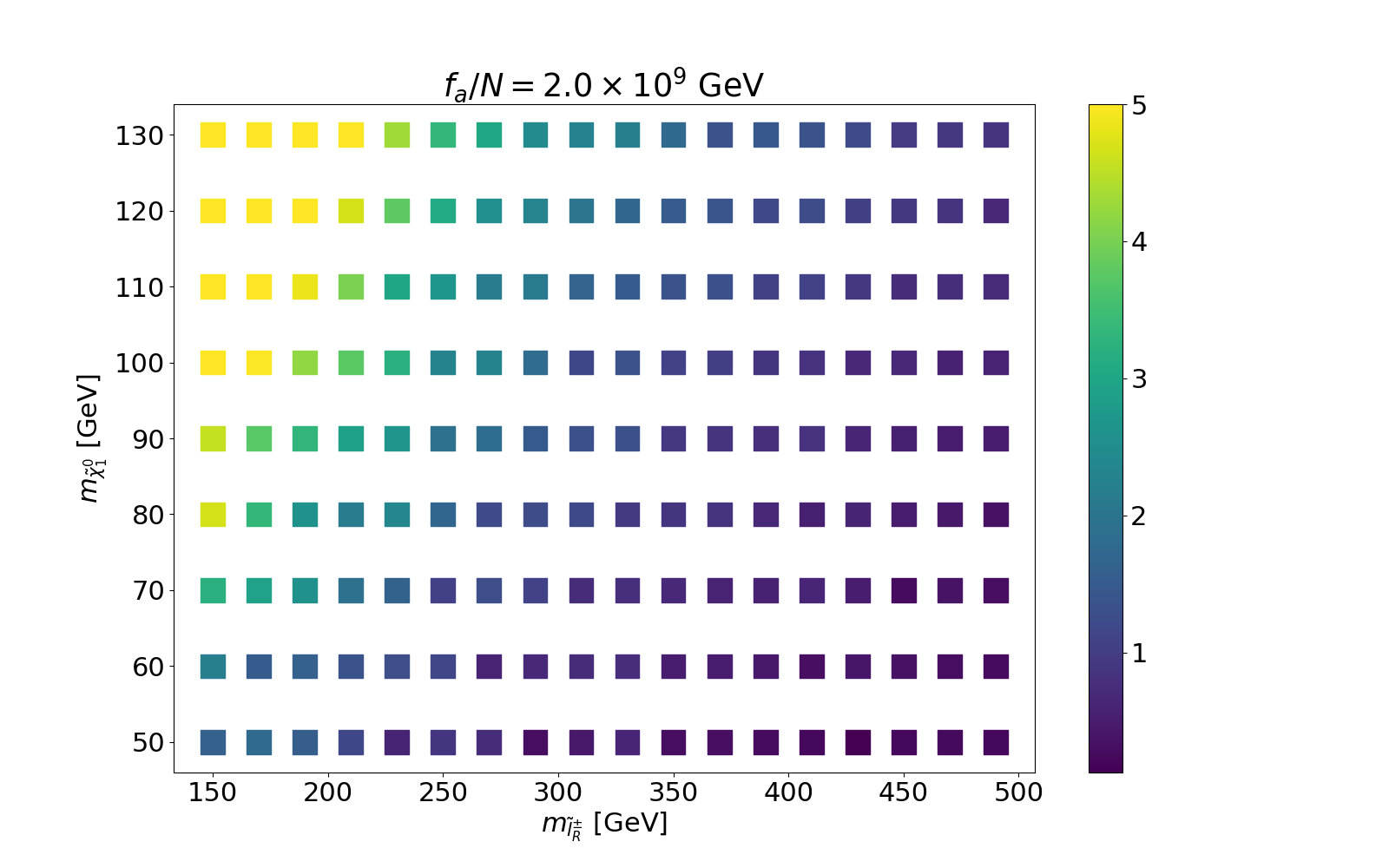}
\includegraphics[width=0.45\textwidth]{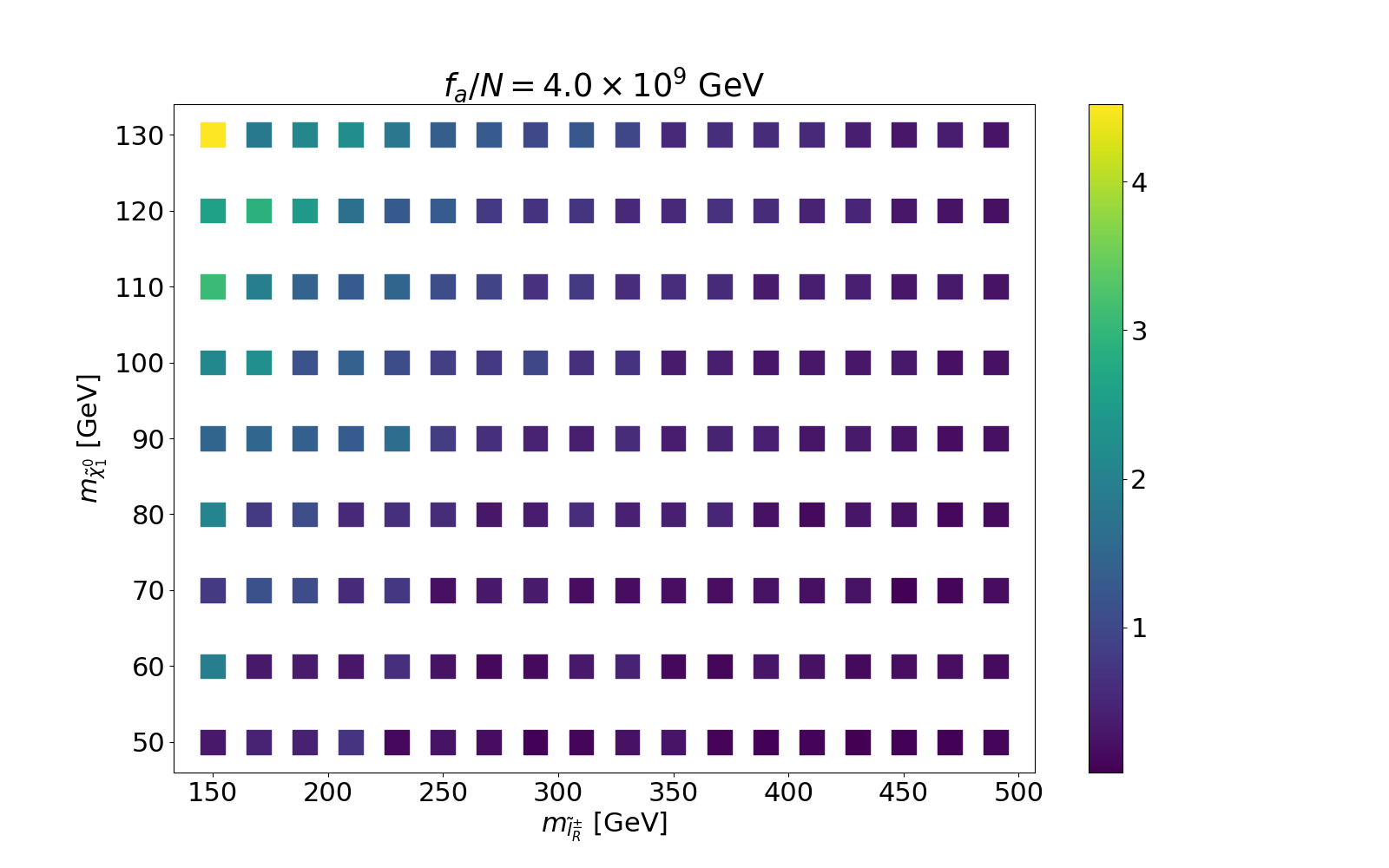}
\includegraphics[width=0.45\textwidth]{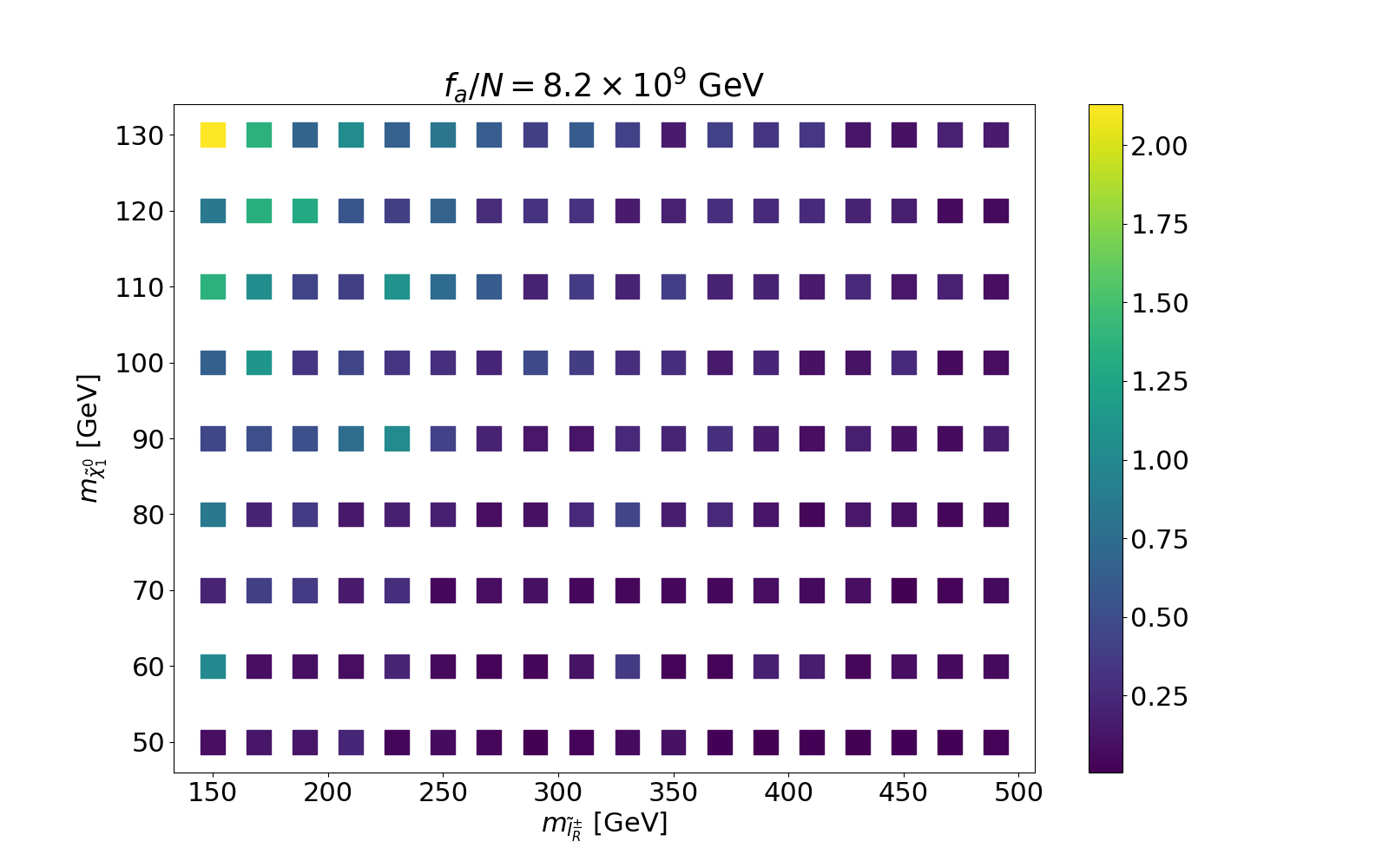}
    \caption {The signal significance for the axino $\tilde{a}$ to be detected by the ECAL.}
\label{fig:collider_result}
\end{figure*}

{As indicated in Eq.~\ref{eq::dw}, the decay length of $\tilde{\chi}_1^0$ becomes larger as the PQ scale $f_a$ grows. For larger $f_a/N$, the neutralino $\tilde{\chi}_1^0$ is more likely to decay outside the ECAL}.
In the Fig.~\ref{fig:collider_result}, the PQ breaking scale is chosen to be between $f_a/N=1.0\times 10^8$ GeV to $5.0\times 10^9$ GeV. The HL-LHC is able to exclude/discovery most of the parameter space for $f_a/N=1.0\times 10^8$ GeV. For $2.5\times 10^8 ~\text{GeV} \lesssim f_a/N \lesssim 7.5\times 10^8 ~\text{GeV}$, the HL-LHC can exclude the $l_R^\pm$ mass region up to around 300 GeV and the $\tilde{\chi}_1^0$ excluded up to about 100 GeV at 95\% C.L.
For $f_a/N \sim  10^9 ~\text{GeV}$,  the $l_R^\pm$ is excluded around 200 GeV and the $\tilde{\chi}_1^0$ excluded up to about 80 GeV at 95\% C.L.
{As we pointed out above, we consider the DFSZ model as an example.  
For the KSVZ model with $C^{\prime}_{aYY} \neq \frac{8}{3}$, the lower limit on $f_a/N$ will be rescaled by a factor of $v_4^{(1)} C^{\prime}_{aYY}/C_{aYY}$ .
}

Before concluding our paper we also display five benchmark points as examples of bino NLSP. Point 1 and Point 2 represent a scenario for $f_{a}=10^{8}$ GeV. Here we see that the bino NLSP lifetime is at the order of $10^{-8}$ s and $10^{-9}$ s, respectively. This means that the bino NLSP will decay just beyond or within ECAL. In rest of the benchmark points the bino NLSP will decay well beyond the ECAL.

\section{Conclusion}

The supersymmetric PQ models, which solve the strong CP problem and provide viable DM candidates simultaneously, 
are also compelling from the collider phenomenological point of view. 
On the one hand, the inclusion of the axino LSP provides a natural solution to the density problem for 
the bino-like lightest neutralino.
On the other hand, the axino is hard to be produced directly by the SM particle collision due to the coupling suppression $1/f_a$, and the axino is, therefore, usually produced via the SUSY particle decays at colliders. 
If the axion coupling lies between $10^{9-10} ~\text{GeV}\lesssim f_a \lesssim 10^{12}$, the lightest neutralino may become a long-lived particle and hence provide a unique signal with very clean background at hadron colliders.
However, the analysis scheme to search for production of the axino LSP via light slepton decay at hadron colliders is inadequately investigated, whose final states include hard leptons, displaced photons and large missing energy. The lack of hard jets makes the displaced vertex difficult to be reconstructed.
In this paper, we concentrated on the scenario where
 the slepton is a few hundred GeV and the bino-like lightest neutralino mass under or around 100 GeV, 
as well as proposed a new analysis method to search for the signal of the long-lived neutralino.
In the analyses, we considered the DFSZ model as an example, and calculated the possibility that the bino-like NLSP decays inside the ECAL and assume zero background. 
We found that for $f_a/N \sim 10^9$ GeV, the right-handed slepton $l^{\pm}_R$ can be excluded around 200 GeV, and 
the bino-like lightest neutralino $\tilde{\chi}_0^1$ can excluded up to about 80 GeV within 2$\sigma$ deviation.

\section{Acknowledgement}

This research is supported in part by the National
Key Research and Development Program of China Grant No. 2020YFC2201504, by the
Projects No. 11875062, No. 11947302, No. 12047503, and No. 12275333 supported by
the National Natural Science Foundation of China, by the Key Research Program of the
Chinese Academy of Sciences, Grant No. XDPB15, by the Scientific Instrument Developing
Project of the Chinese Academy of Sciences, Grant No. YJKYYQ20190049, and by the
International Partnership Program of Chinese Academy of Sciences for Grand Challenges,
Grant No. 112311KYSB20210012.


\newpage
\bibliography{apssamp}

\end{document}